\documentclass{aastex}

\makeatletter
\let\@dates\relax
\makeatother

\usepackage[utf8x]{inputenc}
\usepackage{graphicx}
\usepackage{tabularx}
\usepackage{subfigure}
\usepackage{color}
\usepackage{multicol}
\usepackage{natbib}
\usepackage{mathbbol}
\usepackage{amssymb}
\usepackage{amsmath}
\newcommand{\be}{\begin{equation}}
\newcommand{\ee}{\end{equation}}

\newcommand{\ba}{\begin{eqnarray}}
\newcommand{\ea}{\end{eqnarray}}

\newcommand{\Bf}{{magnetic field}}

\begin{document}

\title{A search for enhanced very-high-energy gamma-ray emission from the March 2013 Crab Nebula flare}

%\author{
%S.~Archambault\altaffilmark{1},
%M.~Lyutikov\altaffilmark{2},
%A.~O'Faol\'ain de Bhr\'oithe\altaffilmark{3},
%A.~Otte\altaffilmark{4},
%J.~Quinn\altaffilmark{3},
%G.~Richards\altaffilmark{4},
%the VERITAS Collaboration
%}

%\altaffiltext{1}{Physics Department, McGill University, Montreal, QC H3A 2T8, Canada}
%\altaffiltext{2}{Department of Physics, Purdue University, 525 Northwestern Avenue, West Lafayette, IN 47907}
%\altaffiltext{3}{School of Physics, University College Dublin, Belfield, Dublin 4, Ireland}
%\altaffiltext{4}{School of Physics and Center for Relativistic Astrophysics, Georgia Institute of Technology, 837 State Street NW, Atlanta, GA 30332-0430}

\author{
E.~Aliu\altaffilmark{1},
S.~Archambault\altaffilmark{2},
T.~Aune\altaffilmark{3},
W.~Benbow\altaffilmark{4},
K.~Berger\altaffilmark{5},
R.~Bird\altaffilmark{6},
A.~Bouvier\altaffilmark{7},
J.~H.~Buckley\altaffilmark{8},
V.~Bugaev\altaffilmark{8},
K.~Byrum\altaffilmark{9},
M.~Cerruti\altaffilmark{4},
X.~Chen\altaffilmark{10,11},
L.~Ciupik\altaffilmark{12},
M.~P.~Connolly\altaffilmark{13},
W.~Cui\altaffilmark{14},
J.~Dumm\altaffilmark{15},
M.~Errando\altaffilmark{1},
A.~Falcone\altaffilmark{16},
S.~Federici\altaffilmark{11,10},
Q.~Feng\altaffilmark{14},
J.~P.~Finley\altaffilmark{14},
P.~Fortin\altaffilmark{4},
L.~Fortson\altaffilmark{15},
A.~Furniss\altaffilmark{7},
N.~Galante\altaffilmark{4},
G.~H.~Gillanders\altaffilmark{13},
S.~Griffin\altaffilmark{2},
S.~T.~Griffiths\altaffilmark{17},
J.~Grube\altaffilmark{12},
G.~Gyuk\altaffilmark{12},
D.~Hanna\altaffilmark{2},
J.~Holder\altaffilmark{5},
G.~Hughes\altaffilmark{11},
T.~B.~Humensky\altaffilmark{18},
P.~Kaaret\altaffilmark{17},
M.~Kertzman\altaffilmark{19},
Y.~Khassen\altaffilmark{6},
D.~Kieda\altaffilmark{20},
F.~Krennrich\altaffilmark{21},
S.~Kumar\altaffilmark{5},
M.~J.~Lang\altaffilmark{13},
M.~Lyutikov\altaffilmark{14},
G.~Maier\altaffilmark{11},
S.~McArthur\altaffilmark{22},
A.~McCann\altaffilmark{23},
K.~Meagher\altaffilmark{24},
J.~Millis\altaffilmark{25,25},
P.~Moriarty\altaffilmark{26},
R.~Mukherjee\altaffilmark{1},
A.~O'Faol\'{a}in de Bhr\'{o}ithe\altaffilmark{6},
R.~A.~Ong\altaffilmark{3},
A.~N.~Otte\altaffilmark{24},
N.~Park\altaffilmark{22},
J.~S.~Perkins\altaffilmark{27},
M.~Pohl\altaffilmark{10,11},
A.~Popkow\altaffilmark{3},
J.~Quinn\altaffilmark{6},
K.~Ragan\altaffilmark{2},
J.~Rajotte\altaffilmark{2},
L.~C.~Reyes\altaffilmark{28},
P.~T.~Reynolds\altaffilmark{29},
G.~T.~Richards\altaffilmark{24},
E.~Roache\altaffilmark{4},
G.~H.~Sembroski\altaffilmark{14},
F.~Sheidaei\altaffilmark{20},
A.~W.~Smith\altaffilmark{20},
D.~Staszak\altaffilmark{2},
I.~Telezhinsky\altaffilmark{10,11},
M.~Theiling\altaffilmark{14},
J.~V.~Tucci\altaffilmark{14},
J.~Tyler\altaffilmark{2},
A.~Varlotta\altaffilmark{14},
S.~P.~Wakely\altaffilmark{22},
T.~C.~Weekes\altaffilmark{4},
A.~Weinstein\altaffilmark{21},
R.~Welsing\altaffilmark{11},
D.~A.~Williams\altaffilmark{7},
A.~Zajczyk\altaffilmark{8},
B.~Zitzer\altaffilmark{9}
}

\altaffiltext{1}{Department of Physics and Astronomy, Barnard College, Columbia University, NY 10027, USA}
\altaffiltext{2}{Physics Department, McGill University, Montreal, QC H3A 2T8, Canada}
\altaffiltext{3}{Department of Physics and Astronomy, University of California, Los Angeles, CA 90095, USA}
\altaffiltext{4}{Fred Lawrence Whipple Observatory, Harvard-Smithsonian Center for Astrophysics, Amado, AZ 85645, USA}
\altaffiltext{5}{Department of Physics and Astronomy and the Bartol Research Institute, University of Delaware, Newark, DE 19716, USA}
\altaffiltext{6}{School of Physics, University College Dublin, Belfield, Dublin 4, Ireland}
\altaffiltext{7}{Santa Cruz Institute for Particle Physics and Department of Physics, University of California, Santa Cruz, CA 95064, USA}
\altaffiltext{8}{Department of Physics, Washington University, St. Louis, MO 63130, USA}
\altaffiltext{9}{Argonne National Laboratory, 9700 S. Cass Avenue, Argonne, IL 60439, USA}
\altaffiltext{10}{Institute of Physics and Astronomy, University of Potsdam, 14476 Potsdam-Golm, Germany}
\altaffiltext{11}{DESY, Platanenallee 6, 15738 Zeuthen, Germany}
\altaffiltext{12}{Astronomy Department, Adler Planetarium and Astronomy Museum, Chicago, IL 60605, USA}
\altaffiltext{13}{School of Physics, National University of Ireland Galway, University Road, Galway, Ireland}
\altaffiltext{14}{Department of Physics, Purdue University, West Lafayette, IN 47907, USA }
\altaffiltext{15}{School of Physics and Astronomy, University of Minnesota, Minneapolis, MN 55455, USA}
\altaffiltext{16}{Department of Astronomy and Astrophysics, 525 Davey Lab, Pennsylvania State University, University Park, PA 16802, USA}
\altaffiltext{17}{Department of Physics and Astronomy, University of Iowa, Van Allen Hall, Iowa City, IA 52242, USA}
\altaffiltext{18}{Physics Department, Columbia University, New York, NY 10027, USA}
\altaffiltext{19}{Department of Physics and Astronomy, DePauw University, Greencastle, IN 46135-0037, USA}
\altaffiltext{20}{Department of Physics and Astronomy, University of Utah, Salt Lake City, UT 84112, USA}
\altaffiltext{21}{Department of Physics and Astronomy, Iowa State University, Ames, IA 50011, USA}
\altaffiltext{22}{Enrico Fermi Institute, University of Chicago, Chicago, IL 60637, USA}
\altaffiltext{23}{Kavli Institute for Cosmological Physics, University of Chicago, Chicago, IL 60637, USA}
\altaffiltext{24}{School of Physics and Center for Relativistic Astrophysics, Georgia Institute of Technology, 837 State Street NW, Atlanta, GA 30332-0430}
\altaffiltext{25}{Department of Physics, Anderson University, 1100 East 5th Street, Anderson, IN 46012}
\altaffiltext{26}{Department of Life and Physical Sciences, Galway-Mayo Institute of Technology, Dublin Road, Galway, Ireland}
\altaffiltext{27}{N.A.S.A./Goddard Space-Flight Center, Code 661, Greenbelt, MD 20771, USA}
\altaffiltext{28}{Physics Department, California Polytechnic State University, San Luis Obispo, CA 94307, USA}
\altaffiltext{29}{Department of Applied Physics and Instrumentation, Cork Institute of Technology, Bishopstown, Cork, Ireland}

\begin{abstract}

In March 2013, a flaring episode from the Crab Nebula lasting $\sim2$ weeks was detected by the \emph{Fermi}-LAT 
(Large Area Telescope on board the \emph{Fermi} Gamma-ray Space Telescope). 
VERITAS provides simultaneous observations throughout this period.
During the flare, the \emph{Fermi}-LAT detected a 20-fold increase in flux above the average synchrotron flux $>100$\,MeV seen
 from the Crab Nebula. Simultaneous measurements with VERITAS are consistent with the non-variable long-term average
 Crab Nebula flux at TeV energies. Assuming a linear correlation between the very-high-energy flux change $> 1$\,TeV
 and the flux change seen in the {\it Fermi}-LAT band $> 100$\,MeV during the period of simultaneous observations, 
 the linear correlation factor can be constrained to be at most $8.6 \times 10^{-3}$ with 95\% confidence.

\end{abstract}


\section{Introduction}
The Crab Nebula is one of the best-studied cosmic particle accelerators. Its distance of $\sim2$\,kpc
 and absolute luminosity of $5\times10^{38}$\,erg s$^{-1}$ allow the study of the nebula in great detail
 across the entire electromagnetic spectrum. From radio to GeV energies, the emission is consistent
 with synchrotron emission of relativistic electrons~\citep{2008ARA&A..46..127H}. However, at higher
 energies, the dominant emission mechanism is thought to be inverse-Compton upscattering of low-energy
 photons by the same population of electrons
~\citep{1965PhRvL..15..577G,1989ApJ...342..379W,1992ApJ...396..161D,2004ApJ...614..897A}. 

The energy source powering the nebula is believed to be the Crab pulsar located at its
 center~\citep{1968Sci...162.1481S}. With the pulsar as the central engine, a self-consistent
 magnetohydrodynamic model can be developed that explains the main features of the
 nebula~\citep{1974MNRAS.167....1R,1984ApJ...283..710K}. The discovery of flaring episodes by the
 AGILE~\citep{2011Sci...331..736T} and {\it Fermi}-LAT~\citep{2011Sci...331..739A} teams was
 unexpected in this framework. The Crab Nebula flux was seen to increase by more than a factor
 of ten in less than a day between 100\,MeV and 1\,GeV in the most extreme of these flares.

Determining the cause of these flares is a major experimental and theoretical challenge. The observed flaring
 timescales of 12 hours~\citep{2011A&A...527L...4B} and 8 hours~\citep{2012ApJ...749...26B} imply that the
 emission region is less than $3\times10^{-4}$\,pc in diameter. This size constraint coupled with the 
observation that the emitted isotropic power peaks at about 1\% of the pulsar
 spin-down power argues in favor of an emission region that moves mildly relativistically
~\citep{2012ApJ...749...26B,2012MNRAS.426.1374C,2012MNRAS.422.3118L,2011MNRAS.414.2229B}.  
 As no enhancement of the pulsed emission has been observed during flares, it has 
 been concluded that the emission region likely resides outside the corotating magnetosphere
~\citep{2012ApJ...749...26B,2011A&A...527L...4B}.

The investigation of the origin of the flares is complicated because no correlated enhancements have been
 observed at other wavelengths to date~\citep{2011A&A...527L...4B,2011ApJ...741L...5S,2013ApJ...765...52S,2012ApJ...749...26B}.
 Multiwavelength campaigns have been executed every time a flare has
 been observed since the detection of the September 2010 flare~\citep{2011Sci...331..736T}. Extensive
 simultaneous coverage over the entire synchrotron emission from radio to X-rays did not reveal correlated
 activity~\citep{2010ATel.3058....1H,2012arXiv1211.7109W} that could have shed light on the location of the flares due
 to better angular resolution at these energies.
 
The non-detection of correlated activity favors a monoenergetic population of relativistic electrons as the
 origin of the observed flares. While multiwavelength coverage has been excellent in radio, optical, and X-rays,
 it has been rather sparse at energies above 100\,GeV, i.e., in the inverse-Compton component. No enhancement of
 the TeV emission was reported by MAGIC or VERITAS during the September 2010 flare~\citep{2010ATel.2967....1M,2010ATel.2968....1O}.
 The ARGO-YBJ Collaboration have reported enhanced signals with a median energy of 1\,TeV from the direction of the
 Crab Nebula contemporaneous to GeV-band flares, although these enhancements did not reach the $5\sigma$ 
 level~\citep{2010ATel.2921....1A,2012ATel.4258....1B,2013arXiv1307.7041V}.

The electrons responsible for the flares
 should also upscatter soft photons in the nebula to produce TeV photons, which enables constraining the dynamics of the electrons. 
In this paper we present the most sensitive observations at TeV energies performed
 during a flare of the Crab Nebula to date. These observations with VERITAS are
 discussed in the context of observations with the {\it Fermi}-LAT.  

\section{Observations and analysis}
\subsection{VERITAS} \label{sec:vts}

The Very Energetic Radiation Imaging Telescope Array System (VERITAS) is an array of four 12\,m diameter
 imaging atmospheric Cherenkov telescopes (IACTs) located at the base of Mt. Hopkins in southern
 Arizona, USA that observes very-high-energy (VHE; $E>100$\,GeV) gamma rays. Each telescope in the array has a
 reflector that is composed of 345 hexagonal mirror facets that  focus light onto a 499-pixel photomultiplier
 tube (PMT) camera at the focal plane with a field of view (FoV) of $\sim3.\!^\circ5$. The array operates in
 the energy range $\sim0.1-30$\,TeV with an energy resolution of $\sim15\%$ at  energies above 1\,TeV and an
 angular resolution of $0.\!^\circ15$~\citep{2008AIPC.1085..657H}.

VERITAS observations of the Crab Nebula in its flaring state were triggered by an automated \textit{Fermi}-LAT
 analysis pipeline at Barnard College-Columbia University~\citep{2011ICRC....8..135E} on 2013 March 02, two days prior to the ATel
 from the \textit{Fermi}-LAT collaboration announcing the gamma-ray flare~\citep{2013ATel.4855....1O}. The VERITAS
 data during the flare are composed of ten nights of observations in the period MJD 56353 to 56366 (2013 March 02
 to 2013 March 15, henceforth referred to as the flare time window, FTW). Observations of the Crab Nebula as part
 of the standard observing schedule from 2012 October 13 to 2013 April 02 excluding the FTW comprise a data set on the source in its
 non-flaring state, which is used as a baseline with which to compare the flare data.

All VERITAS Crab observations were taken in \textit{wobble mode} with an offset of $0.\!^\circ5$ from the source
 position alternately in each of the four cardinal directions, so that the background can be estimated from simultaneously
 gathered data, and systematic effects in the background estimation cancel out~\citep{2001A&A...370..112A, 2007A&A...466.1219B}.
 Observations were conducted using the full four-telescope array in a range of zenith angles $12^\circ - 55^\circ$, giving a
 total of 10.3 hours of live time on the source during the FTW and 17.4 hours during the rest of the season. Two nights of
 flare observations (MJD 56353 and  56358) were conducted at large zenith angles, which has the effect of increasing the 
 effective energy threshold of the array.  Due to this dependence of the energy threshold, the low-energy threshold for
 the spectral analysis is set to a common value of 1\,TeV.  

The recorded images are first flat-fielded using information from nightly calibration runs taken with a pulsed UV 
LED~\citep{2010NIMPA.612..278H}. The images are cleaned using a form of the picture/boundary method~\citep{2008ICRC....3.1325D} 
and parameterized~\citep{1985ICRC....3..445H} to suppress the cosmic ray background. The
 shower direction is reconstructed from the data in each telescope, and a set of selection criteria is applied to
 reject background events~\citep{2001AIPC..558..569K,2008ICRC....3.1325D}.
 
\begin{figure}[ht]
 \centering
 \includegraphics[width=\textwidth]{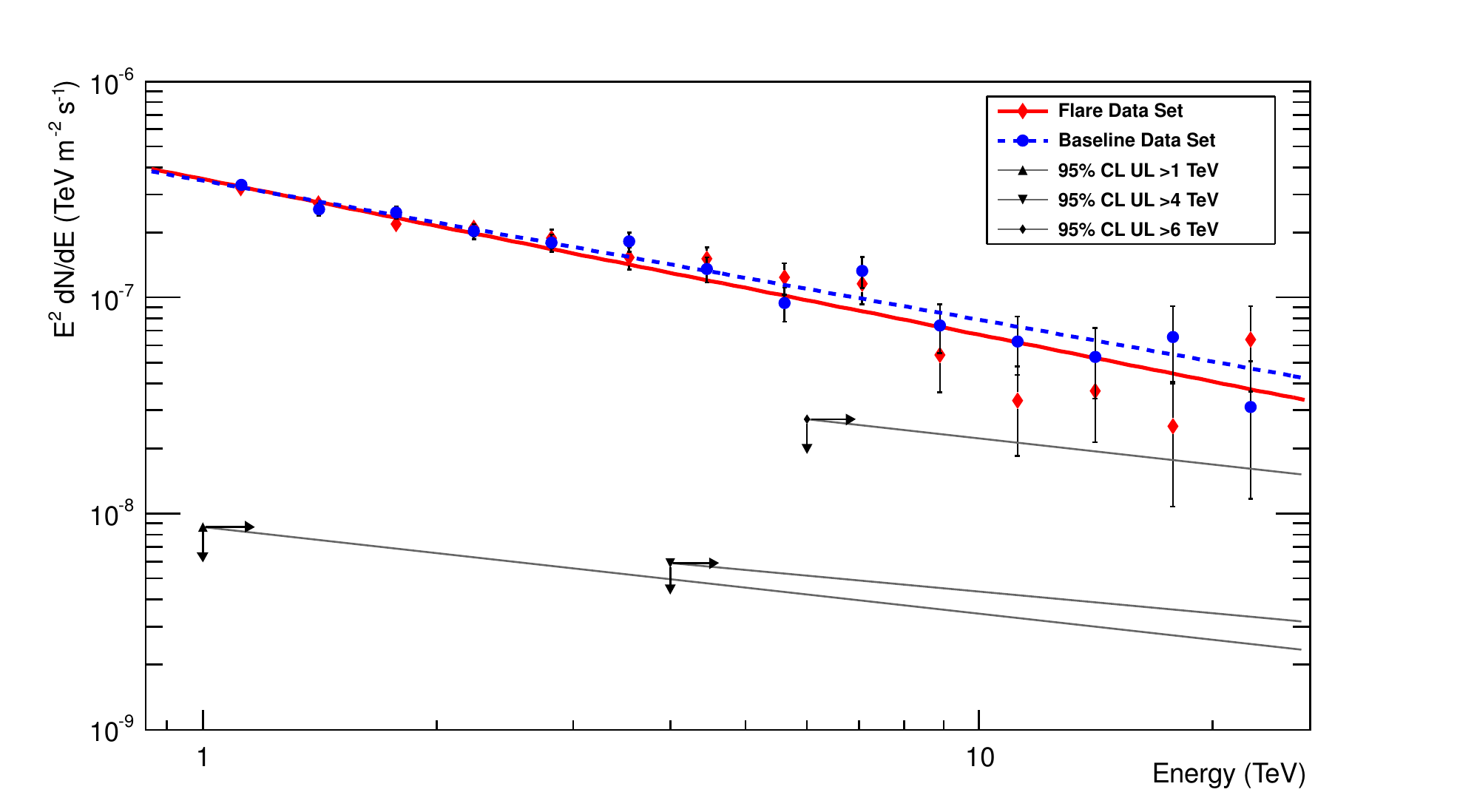}
 \caption{VHE Crab Nebula spectral energy distributions for the flare and non-flare data sets. The SEDs are fit with
 power-law functions (\S~\ref{sec:vts}). From the limits on the relative flux change above 1\,TeV, 4\,TeV, and 6\,TeV
 (\S~\ref{sec:results}), upper limits on an extra flux component in the flare are computed assuming a spectral index of $-2.4$.}
 \label{fig:sed}
 \end{figure}

Energy spectra are calculated $> 1$\,TeV both for the FTW and the baseline observations and are shown in Figure~\ref{fig:sed}.
 The spectra are parameterized as power laws of the form 
 \begin{equation}
 {dN \over dE} = N_0 \left({E \over \textrm{1\,TeV}} \right)^{\gamma}.
 \end{equation}
 The baseline spectral fit gives a normalization of $N_0^{\textrm{baseline}} = (3.48 \pm 0.14_{\textrm{stat.}} \pm 1.08_{\textrm{sys.}}) 
\times 10^{-7} \ \textrm{TeV}^{-1} \ \textrm{m}^{-2} \ \textrm{s}^{-1}$ and $\gamma^{\textrm{baseline}} = -2.65 
\pm 0.04_{\textrm{stat.}} \pm 0.3_{\textrm{sys.}}$,
 with a $\chi^2$ value of 16.6 with 12 degrees of freedom (dof). The FTW spectral fit gives a normalization of
 $N_0^{\textrm{flare}} = (3.53 \pm 0.15_{\textrm{stat.}} \pm 1.12_{\textrm{sys.}}) \times 10^{-7} \ \textrm{TeV}^{-1} \ \textrm{m}^{-2} \ \textrm{s}^{-1}$ and
 a spectral index $\gamma^{\textrm{flare}} = -2.72 \pm 0.05_{\textrm{sys.}} \pm 0.3_{\textrm{sys.}}$, with a $\chi^2$ value of 10.1 with 12 dof. The fit
 probabilities are 16\% and 61\%, respectively. These spectral parameters are mutually consistent, implying no change
 of the TeV flux during the FTW.  The systematic uncertainties on the flux normalization and spectral index are expected to vary slowly with time, and a paper 
containing a proper treatment of these uncertainties is currently in preparation.

\subsection{{\it Fermi}-LAT} \label{sec:lat}
The {\it Fermi}-LAT is a pair-conversion telescope sensitive to gamma-ray photons with energies between 20\,MeV and
 300\,GeV.  It has a wide FoV of $\sim2.5$\,sr and surveys the entire sky every three hours.  For a complete description of the
 instrument, see~\cite{2009ApJ...697.1071A,2012ApJS..203....4A}.

In order to extract spectral parameters of the Crab, the {\it Fermi}-LAT \texttt{Science Tools v9r27p1}
 with \texttt{P7V6} instrument response functions (IRFs) and the standard quality cuts described in~\cite{2012ApJS..199...31N}
 are used. Two years of ``source''-class events with energies between 100\,MeV and 300\,GeV collected 
 between MJD 54832 and 55562 within $20^\circ$ of the Crab are processed with the maximum likelihood fitting routine. A model
 of the background is obtained in a binned likelihood analysis by fitting spectral models for all sources in the 2FGL catalog
 within $20^\circ$ of the Crab in addition to the galactic and isotropic diffuse backgrounds (\texttt{gal\_2yearp7v6\_v0.fits},
 \texttt{iso\_p7v6source.txt}). Photon arrival times are barycentered with \texttt{Tempo2}~\citep{2006MNRAS.369..655H} using
 a publicly-available Jodrell Bank radio ephemeris for the Crab pulsar~\citep{1993MNRAS.265.1003L} to allow a selection of
 the off-pulse phase region $0.48 - 0.88$. Under the assumption that emission from the pulsar is negligible in the off-pulse
 region, spectral parameters for the synchrotron and inverse-Compton components of the Crab Nebula are calculated. These
 parameters are fixed in the model to allow fitting of the pulsar spectral component after undoing the
 selection on pulsar phase.

The Crab Nebula synchrotron differential spectrum is parameterized as a power law (PowerLaw2 in the
 {\it Fermi}-LAT \texttt{Science Tools}) of the form
 \begin{equation}
 {dN \over dE} = {F_0(\gamma + 1)E^{\gamma} \over (\textrm{300\,GeV})^{\gamma + 1} - (\textrm{0.1\,GeV})^{\gamma + 1}}.
 \end{equation}
 The fit of the quiescent state yields a synchrotron integral flux above 100\,MeV of
 $F_0 = (6.40 \pm 0.11) \times 10^{-7} \ \textrm{cm}^{-2} \ \textrm{s}^{-1}$ and photon index of
 $\gamma = -3.69 \pm 0.11$, which are consistent with previously published results~\citep[e.g.,][]{2012ApJ...749...26B}. 

\begin{figure}[t]
 \centering
 \includegraphics[width=\textwidth]{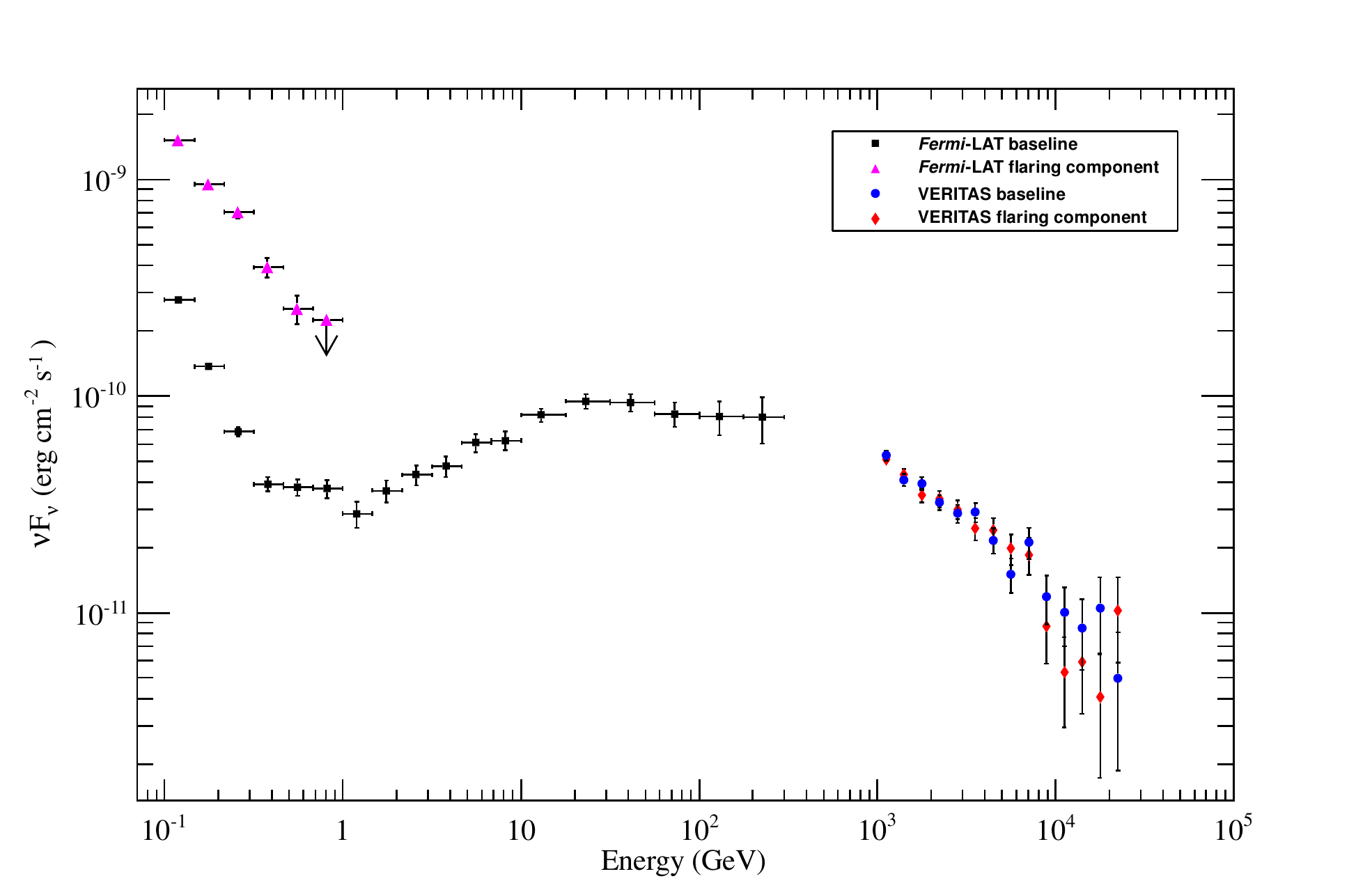}
 \caption{Combined SED of the Crab Nebula. The baseline {\it Fermi}-LAT spectrum (black squares) is
 averaged over $\sim5$ years of observations, while the baseline VHE spectrum (blue circles) includes all good data taken
 outside of the FTW in the $2012-2013$ VERITAS observing season. The FTW VHE spectrum (red diamonds) shows no significant
 deviation from the baseline, while the synchrotron spectrum during this period (magenta triangles) exhibits spectral
 hardening. All spectral parameters given in \S~\ref{sec:vts},~\ref{sec:lat}.}
 \label{fig:combined_sed}
 \end{figure}

A similar analysis is done for the FTW. Since the {\it Fermi}-LAT carried out a targeted observation of the Crab
 during the flare, the recommended \texttt{P7V6MC} IRFs and pointed mode data selection criteria are used in this 
analysis\footnote{\footnotesize
 http://fermi.gsfc.nasa.gov/ssc/data/analysis/documentation/Cicerone/Cicerone\_Likelihood/Exposure.html}.  
 The synchrotron integral flux above 100\,MeV for the FTW is found to be
 $(5.30 \pm 0.13) \times 10^{-6} \ \textrm{cm}^{-2} \ \textrm{s}^{-1}$ with a harder photon index of $-3.10 \pm 0.05$.
 A combined spectral energy distribution (SED) showing the {\it Fermi}-LAT and VERITAS spectra is given in Figure~\ref{fig:combined_sed}.

\section{Results} \label{sec:results}
\begin{figure}[ht]
 \centering
 \includegraphics[width=\textwidth]{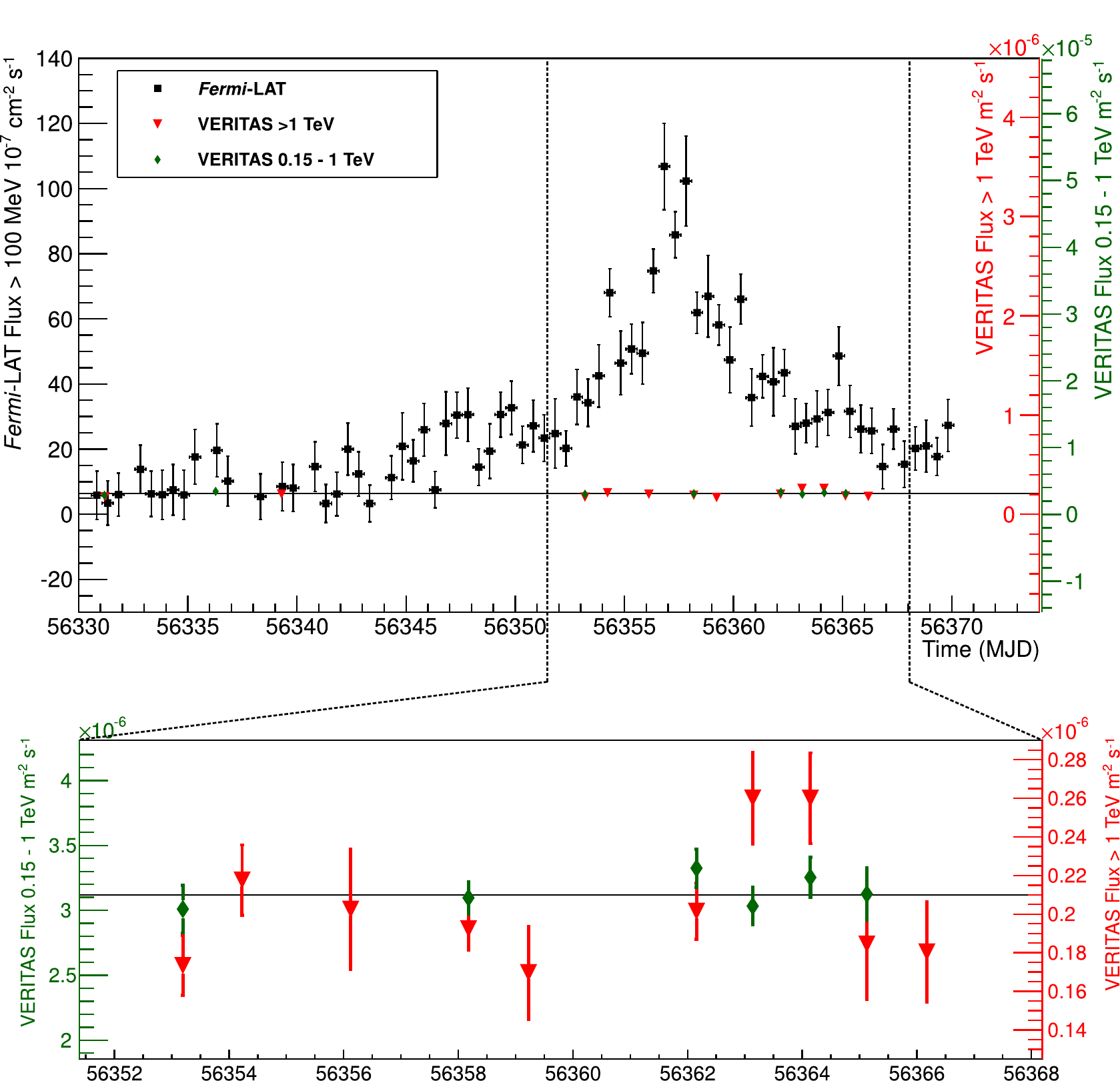}
 \caption{{\it Fermi}-LAT and VERITAS light curves for the March 2013 Crab Nebula flare. The 12-hour binned {\it Fermi}-LAT
 light curve (square markers) spans MJD $56330 - 56370$. The VERITAS light curves (triangle and diamond markers) span ten nights during the
 FTW where weather permitted observations. The baseline Crab Nebula synchrotron flux above 100\,MeV and average VHE flux above
 0.15\,TeV and 1\,TeV are aligned and are indicated by the solid black line. The vertical scales of the three light curves have been adjusted such
 that the zero points and baseline fluxes are coincident.}
 \label{fig:light_curves}
 \end{figure}

A test for variability in the VERITAS FTW light curve (shown in Figure~\ref{fig:light_curves}) is performed by
 fitting the light curve with a constant flux. This fit gives a flux $> 1$\,TeV of
 $(2.05 \pm 0.07) \times10^{-7} \ \textrm{m}^{-2} \ \textrm{s}^{-1}$  with a $\chi^2$ value of 19.1 with
 9 dof (probability $\sim2.4\%$). 
By fitting a light curve of data taken outside of the FTW, the Crab Nebula is detected with a baseline VHE flux 
$> 1$\,TeV of $(2.10 \pm 0.06) \times 10^{-7} \ \textrm{m}^{-2} \ \textrm{s}^{-1}$ with a $\chi^2$ value of 21.7
 with 22 dof (probability $\sim47.8$\%). The FTW flux is thus consistent with the baseline flux and with
 no statistically significant variability during the flare.  An analysis of a subset of the data with energies extending down to 
$\sim150$ GeV was also conducted (shown in Figure~\ref{fig:light_curves}), however no variability is revealed at these energies.

In order to test for correlated {\it Fermi}-LAT and VERITAS ($> 1$\,TeV) flux variability in the light curves shown in
 Figure~\ref{fig:light_curves}, a publicly available implementation of the $z$-transformed discrete correlation function
 (ZDCF) is employed~\citep{1997ASSL..218..163A,2013arXiv1302.1508A}.
 The ZDCF method requires a minimum of 12 observations in each light curve for a statistically valid analysis, so
 two nights of pre-flare VERITAS Crab Nebula observations taken on MJD 56331 and 56339 (February 8 and 16, respectively)
 are added before the cross-correlation is performed. The zero time-lag bin reported a ZDCF correlation coefficient of
 \begin{equation}
 \textrm{DCF} = -0.07 \pm 0.31
 \end{equation}
 which is consistent with no correlation at zero lag. Results for all other time-lag bins are also consistent with no statistically
 significant correlation.

\begin{figure}[t]
 \centering
 \includegraphics[width=\textwidth]{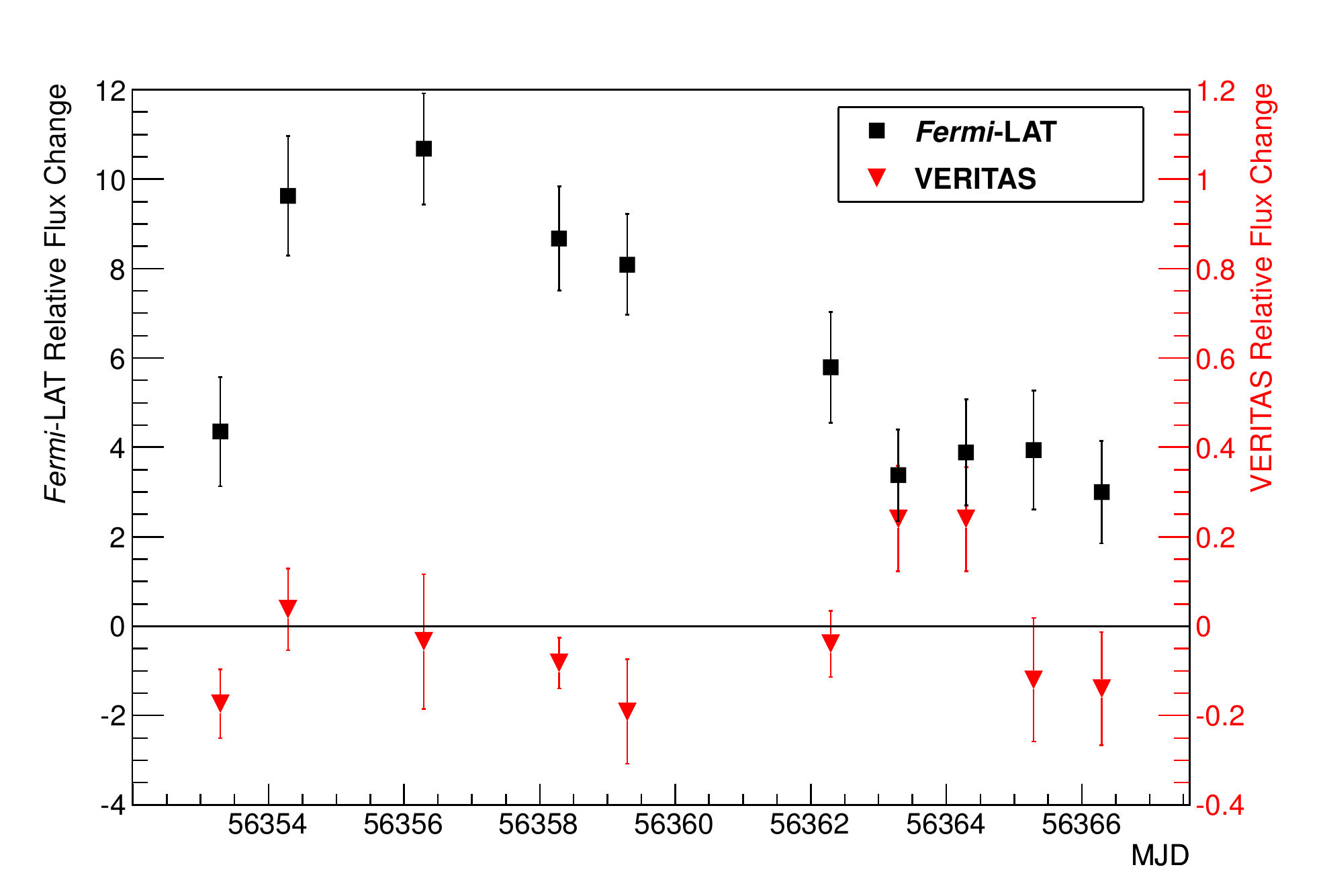}
 \caption{Relative flux changes for simultaneous {\it Fermi}-LAT (square markers) and VERITAS (triangle markers) observations during the FTW.
 The zero line corresponds to an observed flux equal to the average. Note that the vertical scale for the VERITAS points is a factor of ten
 smaller than the vertical scale for the {\it Fermi}-LAT points.}
 \label{fig:rel_flux_changes}
\end{figure}

Relative flux changes during the FTW are calculated for VERITAS and {\it Fermi}-LAT.  The $i^{\textrm{th}}$ relative flux change
 $\Delta F_{\textrm{rel.}}^{i}$ for both VERITAS and {\it Fermi}-LAT observations on the $i^{\textrm{th}}$ night is computed as
 \begin{equation}
 \Delta F_{\textrm{rel.}}^{i} = \frac{ {F^i - \overline{F}}}{\overline{F}}.
 \end{equation}
 For VERITAS, $F^i$ is the average flux for one night. For {\it Fermi}-LAT, $F^i$ is the average flux in one 12-hour time bin
 centered on midnight Arizona time (MST, 0700 UTC). $\overline{F}$ is the average non-flare flux from the nebula. The VERITAS and
 {\it Fermi}-LAT  relative flux changes for simultaneous observations are shown in Figure~\ref{fig:rel_flux_changes}.
 Averaged over the simultaneous observations in the FTW, the relative flux changes are
\begin{align} 
  &\overline{\Delta F^{\textrm{VTS}}_{\textrm{rel.}}} = -0.026 \pm 0.035 \ \ &\textrm{(VERITAS} > \textrm{1\,TeV)}         \\
  &\overline{\Delta F^{\textrm{Fermi}}_{\textrm{rel.}}} = 6.14 \pm 0.38  \ \ &\textrm{({\it Fermi}-LAT} > \textrm{100\,MeV)}
\end{align}

From $\overline{\Delta F^{\textrm{VTS}}_{\textrm{rel.}}}$, a 95\% confidence level upper limit (UL) is computed for an elevated VHE
 flux. Given the assumption of a positive and non-zero correlation of {\it Fermi}-LAT and VERITAS flux changes, a Bayesian prior
 is introduced in the limit calculation that is zero for negative relative flux changes and one elsewhere. This prior is equivalent
 to invoking the physical constraint that all of the VHE flux changes are at least zero. The upper limit is calculated over
 the Bayesian interval $[0,x^{\textrm{up}}]$ such that  
 \begin{equation}
  {\frac{ \int\limits_0^{x^{\textrm{up}}} \textrm{exp}\left({-\frac{(\overline{\Delta F^{\textrm{VTS}}_{\textrm{rel}}}-x)^2}{2\sigma^2}}\right)dx}{ \int\limits_0^{\infty} 
\textrm{exp}\left({-\frac{(\overline{\Delta F^{\textrm{VTS}}_{\textrm{rel}}}-x')^2}{2\sigma^2}}\right)dx'}} = 0.95 \ \ (x^{\textrm{up}} > 0)
 \end{equation}
 where $\sigma$ is the error on $\overline{\Delta F^{\textrm{VTS}}_{\textrm{rel.}}}$, and the $95\%$ CL upper limit
 is given by $x^{\textrm{up}}$, which is obtained by solving the equation numerically. Limits are calculated for
 three different energy thresholds shown in Table~\ref{tab:energybandtable}.

\newcolumntype{C}{>{\centering\arraybackslash}X}
\begin{table}[ht!]
\centering
\begin{tabularx}{0.75\textwidth}{CCC}
  \hline
  Energy band (TeV) & $\overline{\Delta F^{\textrm{VTS}}_{\textrm{rel.}}}$ $95\%$ CL UL & $95\%$ CL integral UL
 at threshold (TeV m$^{-2}$ s$^{-1}$) \\ \hline
  $>1$ & $5.3\%$  & $8.7 \times 10^{-9}$ \\ 
  $>4$ & $6.8\%$  & $5.9 \times 10^{-9}$ \\
  $>6$ & $37.4\%$ & $2.7 \times 10^{-8}$ \\
  \hline
\end{tabularx}
\caption{95\% CL Bayesian upper limits on the VHE relative flux increase during the flare
 period for three energy thresholds.}
\label{tab:energybandtable}
\end{table}

By adopting the assumption that the relative flux change seen by VERITAS is linearly related to that seen by the
 {\it Fermi}-LAT during the flare:
 \begin{equation}
 \overline{\Delta F^{\textrm{VTS}}_{\textrm{rel.}}} = \alpha \overline{\Delta F^{\textrm{{\it 
Fermi}}}_{\textrm{rel.}}},
 \end{equation}
 a constraint on the linear correlation factor $\alpha$ can be calculated, which can be used to test model predictions.  
Taking the ratio of the $> 1$\,TeV upper limit and
 the average {\it Fermi}-LAT relative flux change, we find that $\alpha < 8.6 \times 10^{-3} \ \textrm{(95\% CL)}$ for the average
 of the ten nights of simultaneous observations. The constraint on $\alpha$ is also computed night-by-night, though only
 MJD 56358 gives the slightly better constraint of $\alpha < 8.1 \times 10^{-3} \ \textrm{(95\% CL)}$.

\section{Discussion}
In this paper we present observations of the Crab Nebula with VERITAS and the {\it Fermi}-LAT during the March 2013 flare. The light
 curve and reconstructed energy spectrum between 1\,TeV and 10\,TeV do not indicate any flux enhancement at TeV energies, while
 the flux above 100\,MeV was six times elevated during our observations.

Earlier flares had very hard spectra with peak energy reaching up to $\epsilon_\mathrm{flare}\approx 500$\,MeV~\citep{2012ApJ...749...26B},
though in the present flare, a peak could not be resolved in the MeV -- GeV spectrum leaving the electron spectrum unconstrained
 at lower energies. The synchrotron spectrum above 100\,MeV is slightly harder than for the quiescent Crab, which may reflect
 a separate electron population and/or an increase in the magnetic-field strength in the emission zone that shifts a harder
 section of a curved synchrotron spectrum into the frequency band accessible with the \textit{Fermi}-LAT. Neglecting the weak
 modifications arising from the possibility of mildly relativistic bulk motion, we suggest that some excess electron acceleration
 took place.

 From classical electrodynamics, the Lorentz factor of electrons that would emit 200\,MeV synchrotron radiation is
\begin{equation} 
\gamma_\mathrm{sy}=3\times 10^9\, \left(\frac{B}{\mathrm{mG}}\right)^{-0.5}\ ,  
\label{eq:1} 
\end{equation}
and their energy-loss rate and life time are
\begin{equation} 
\dot E_\mathrm{sy}=(8\times 10^{-3}\ \mathrm{erg/s})\,\left(\frac{B}{\mathrm{mG}}\right),\qquad
\tau_\mathrm{sy}=(3\times 10^5\ \mathrm{s})\,\left(\frac{B}{\mathrm{mG}}\right)^{-1.5}\ .  
\label{eq:2}
\end{equation}
Assuming a magnetic field of 1\,mG in the emission region, similar to that deduced in~\cite{2011MNRAS.414.2229B},
 the flare duration $\tau_\mathrm{sy}$ is on the order of a few days, which is consistent with observed flares
 at a few hundred MeV.  If the \Bf\ were significantly stronger than 1\,mG, the synchrotron lifetime would become
 very short compared to the flare duration, and so the electron population would need to be continuously
 replenished to sustain the flare.  Thus, the main cause of the synchrotron flare was likely the injection of a
 large number of excess electrons at PeV energies.  %Any moderate bulk motion would not change this picture.

\cite{2011MNRAS.414.2229B} consider a model in which electrons are injected into the magnetic field of the pulsar wind
 zone and produce synchrotron gamma rays through acceleration in reconnection regions of the magnetic field. Assuming
 the electrons reach an equilibrium spectrum described by a differential power law with index between $3.0-3.6$ and
 with a characteristic cut-off at $\gamma=3\times10^9$ for flares, they suggest variability above $\sim1$\,TeV of
 roughly $10\%$ with more substantial changes above $\sim10$\,TeV as a result of inverse-Compton scattering. 
 However, inverse-Compton scattering of soft photons by electrons with Lorentz
 factors $\sim10^9$ is heavily Klein-Nishina suppressed and would provide gamma rays in the PeV band, beyond the reach
 of VERITAS. Excess electrons with Lorentz factors of $\gamma_{\textrm{IC}} \simeq10^7$ may produce a flux enhancement at TeV energies, but the
 non-detection with VERITAS poses challenges for this model and thus constrains the number of 
electrons with Lorentz factors of $\gamma_{\textrm{IC}}$.

The number of electrons with Lorentz factors of $\sim3\times 10^9$ can be estimated as
\begin{equation} 
N_\mathrm{e,sy} = \frac{L_\mathrm{sy}}{\dot E_\mathrm{sy}}\simeq 6\times10^{37} 
\,\left(\frac{B}{\mathrm{mG}}\right)^{-1}\ , \label{eq:3} 
\end{equation}
where $L_\mathrm{sy}$ is the synchrotron luminosity at 200\,MeV. 
To calculate the number of electrons that may inverse-Compton scatter soft (infrared, IR) photons into the TeV band, we
 need to know the density of low-frequency radiation in the nebula. To this end we use
 $ L_\mathrm{soft}  \sim10^{37}$\,erg s$^{-1}$~\citep{1984ApJ...278L..29M} as the
 pulsar wind nebula (PWN) luminosity in IR photons, $ \epsilon_\mathrm{soft} \sim0.1$\,eV as the photon energy,
 and $ d_\mathrm{PWN} \simeq1$\,pc as the characteristic size of the Crab Nebula. The density of IR photons is then 
 \be
 n_\mathrm{soft} \simeq \frac{L_\mathrm{soft}}{4 \pi\,
 d_\mathrm{PWN}^2\, c \epsilon_{\textrm{soft}}}\simeq 20\ \mathrm{cm^{-3}}. \label{eq:4}
 \ee
 Using the upper limit on an extra flux component $>1$\,TeV given in Table~\ref{tab:energybandtable}, we find that the
 inverse-Compton luminosity $L_{\textrm{IC}} \lesssim 4 \times 10^{32} \ \textrm{erg} \ \textrm{s}^{-1}$.  The number of
 electrons that upscatter photons to TeV energies is given by
 \be
 N_\mathrm{e,IC} = \frac{L_\mathrm{IC}}{\sigma_\mathrm{T} n_\mathrm{soft} c \times \gamma m_e c^2},
 \ee
 where $\sigma_\mathrm{T}$ is the Thomson cross-section.  Ignoring the moderate Klein-Nishina suppression (the kinematic parameter
 $4\,\epsilon_\mathrm{soft}\epsilon_\gamma / (m_e^2\,c^4)\simeq 10$), the upper limit derived on excess TeV gamma rays
 corresponds to at most
 \be
 N_\mathrm{e,IC} \left(\gamma\approx 10^7\right)\lesssim 10^{44}\ .  \label{eq:5}
 \ee
 Assuming for ease of exposition that the spectrum of excess electrons follows a power law, $N_e (\gamma)\propto \gamma^{-s}$,
 the corresponding constraint on the spectral index is
 \be
 s\lesssim \frac{6.2+\log\left(\frac{B}{\mathrm{mG}}\right)}{2.5-\frac{1}{2}\log\left(\frac{B}{\mathrm{mG}}\right)}\
 , \label{eq:6}
 \ee
 which permits $s\simeq2.5$ for the fiducial magnetic-field strength of 1\,mG.  This index is harder than that assumed 
by ~\cite{2011MNRAS.414.2229B} and constrains the number of electrons that may be responsible for the Crab flare.  Future 
observations with VERITAS or next-generation telescope arrays will likely provide more stringent constraints.  

\acknowledgments
This research is supported by grants from the U.S. Department of Energy Office of Science,
 the U.S. National Science Foundation and the Smithsonian Institution, by NSERC in Canada, by
 Science Foundation Ireland (SFI 10/RFP/AST2748) and by STFC in the U.K. We acknowledge the excellent
 work of the technical support staff at the Fred Lawrence Whipple Observatory and at the collaborating
 institutions in the construction and operation of the instrument.

The authors are grateful to M. Mayer, E. Hays, and R. Buehler for a useful discussion of the {\it Fermi}-LAT 
data.

A. O'Faol\'ain de Bhr\'oithe acknowledges the support of the Irish Research Council ``Embark Initiative''.

N.O. and G.R. gratefully acknowledge the support of a Fermi Guest Investigator grant.

\pagebreak
\newpage
\pagebreak
\newpage

\pagebreak
\newpage

\pagebreak
\newpage

%===============
% Bibliography
%===============
%\bibliographystyle{apj}
%\bibliography{./CrabFlare.bib}

\end{document}